\documentclass[aps,pre,letterpaper,twocolumn,showpacs]{revtex4-1}
\usepackage[utf8]{inputenc}
\usepackage{multirow}
\usepackage{amsmath,amsfonts,amssymb}

\usepackage{graphicx}
\usepackage{psfrag}

\def\be{\begin{equation}}
\def\ee{\end{equation}}
\def\ba{\begin{eqnarray}}
\def\ea{\end{eqnarray}}

\begin{document}
\title{First-order transition and marginal critical behavior\\
in a novel 2D frustrated Ising model}
\author{Christophe Chatelain}
\affiliation{Universit\'e de Lorraine, CNRS, LPCT, F-54000 Nancy, France}
\date{\today}

\begin{abstract}
The phase diagram of a novel two-dimensional frustrated Ising model
with both anti-ferromagnetic and ferromagnetic couplings is studied
using Tensor-Network Renormalization-Group techniques. This model can
be seen as two anti-ferromagnetic Ising replicas coupled by non-local
spin-spin interactions, designed in such a way that the continuum limit
matches that of the still debated $J_1-J_2$ model and induces a
marginal critical behavior. Our model has the advantage of having more
symmetries than the $J_1-J_2$ model and of allowing a more straightforward
implementation of Tensor-Network Renormalization-Group algorithms
We demonstrate the existence
of two transition lines, featuring both first and second-order regimes.
In the latter, the central charge and the critical exponents are shown
to be compatible with the Ashkin-Teller universality class. This picture
is consistent with that given by Monte Carlo simulations of the
$J1-J2$ model but not with recent studies with Tensor-Network techniques.
\end{abstract}
\maketitle

\section{Introduction}
\begin{figure}
    \centering
    \includegraphics[width=0.23\textwidth]{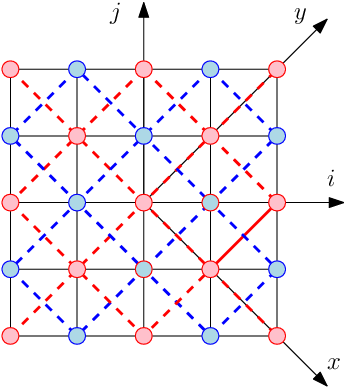}
    \caption{Ising model with nearest and next-to-nearest couplings.
    The black edges are the ferromagnetic couplings $J_1$. The dashed edges
    are anti-ferromagnetic with a coupling $J_2$. The blue and red colors
    correspond to the two sublattices that are decoupled when $J_1=0$.}
    \label{fig1}
\end{figure}

Despite of decades of active research, the phase diagram of the Ising
model with ferromagnetic couplings $J_1$ between nearest-neighboring
spins and anti-ferromagnetic ones $J_2$ between next-nearest ones
remains controversial. On the square lattice, the Hamiltonian of the
$J_1-J_2$ model is
    \begin{eqnarray}
    -\beta H&=&J_1\sum_{i,j}\sigma_{i,j}
    \big[\sigma_{i+1,j}+\sigma_{i,j+1}\big]\nonumber\\
    &&\quad -J_2\sum_{i,j}\big[\sigma_{i,j}\sigma_{i+1,j+1}
    +\sigma_{i+1,j}\sigma_{i,j+1}\big]
    \label{Eq2}
    \end{eqnarray}
with $\sigma_{i,j}\in\{-1;+1\}$.
The ferromagnetic Ising model is recovered when $J_2=0$. The system undergoes
therefore a ferromagnetic-paramagnetic second-order phase transition at the
self-dual coupling $J_1={1\over 2}\ln(1+\sqrt 2)$~\cite{Cardy1,Mussardo}.
The critical behavior belongs to the Ising universality class with magnetic
and thermal critical exponents $\beta=1/8$ and $\nu=1$ and a central charge $c=1/2$.
When $J_1=0$, the system decouples into two sublattices, at $45^\circ$
of the original lattice and with a lattice step $\sqrt 2$ (Fig.~\ref{fig1}).
Each one of these sublattices undergoes a transition from an anti-ferromagnetic
phase to the paramagnetic phase at $J_2={1\over 2}\ln(1+\sqrt 2)$. The
associated critical behavior also belongs to the Ising universality class.
The central charge of the model is $c=2\times{1\over 2}=1$.
\\

For non-zero $J_1$ and $J_2$, the ground state is readily obtained by
minimizing the energy of a plaquette of 4 spins. When $g=J_2/J_1<1/2$,
the ground states are the two ferromagnetic spin configurations. In
contrast, when $g>1/2$, the ground state is ferromagnetic in one direction
of the lattice but anti-ferromagnetic in the other one. The four possible
spin configurations show stripes, either horizontal or vertical.
These ground-states are sometimes referred to as super-antiferromagnetic.
Early transfer-matrix and Monte Carlo simulations suggested the existence
of a line of continuous phase transition to the paramagnetic
phase~\cite{Nightingale,Swendsen,Binder,Oitmaa}.
The critical behavior belongs to the Ising universality class in the
regime $g<1/2$, i.e. for the ferromagnetic-paramagnetic transition.
However, in the regime $g>1/2$, i.e. for the transition between the
super-antiferromagnetic and paramagnetic phases, critical exponents were
observed to vary with the ratio $g$ but the ratio $\beta/\nu$ remains
constant, in agreement with the
weak-universality scenario~\cite{Suzuki}. One decade later, Mor\'an-L\'opez
{\sl et al.} showed by mean-field calculations the existence of a regime
of first-order phase transition when $1/2<g\lesssim 1.14$~\cite{Lopez1,Lopez2}.
Later Monte Carlo simulations confirmed the existence of this regime
but only in the range $1/2<g\lesssim 0.67$~\cite{Kalz1,Jin1,Kalz2,Jin2}.
\\

These studies also agree on the fact that, in the second-order regime
$g\gtrsim 0.67$, the critical behavior belongs to the Ashkin-Teller
universality class. The Ashkin-Teller model consists in two Ising models
coupled by their energy densities. This coupling is marginal and leads to
non-universal critical exponents~\cite{Kadanoff,Nienhuis,Baxter}. As
shown on Fig.~\ref{fig1}, the $J_1-J_2$ Ising model can also be viewed
as two anti-ferromagnetic Ising models that are coupled by a spin-spin interaction,
rather than an energy-energy one. However, this interaction does not couple
spins at the same site, the perturbation would be irrelevant in that case, but
at different sites. Kalz {\sl et al.} argued that this perturbation is marginal,
as in the Ashkin-Teller model~\cite{Kalz1}. Their argument goes as follows.
Introduce the spins on the two sublattices as $\sigma^A_{x,y}$ and
$\sigma^B_{x,y}$ where the lattice coordinates $(x,y)$ shown on
Fig.~\ref{fig1} are integers on the sublattice A and half-integers on B.
The total energy associated to the ferromagnetic couplings between the
two sublattices reads
    \ba &&J_1\sum_{x,y}\sigma^A_{x,y}\big[
    \sigma^B_{x-1/2,y-1/2}+\sigma^B_{x-1/2,y+1/2}\big.\nonumber\\
    &&\quad\quad
    \big.+\sigma^B_{x+1/2,y-1/2}+\sigma^B_{x+1/2,y+1/2}\big].
    \label{Eq1}
    \ea
In the super-antiferromagnetic phase, the two sublattices are
anti-ferromagnetically ordered, so it is convenient to introduce
the staggered magnetization as
    \be S_{x,y}^{A}=(-1)^{x+y}\sigma_{x,y}^A,\quad
    S_{x,y}^{B}=(-1)^{x+y}\sigma_{x,y}^B.\label{Stag}\ee
In the continuum limit, the energy Eq.~\ref{Eq1} tends to
    \ba &&J_1\sum_{x,y}S^A_{x,y}
    \Big[-S^B_{x-1/2,y-1/2}+S^B_{x-1/2,y+1/2}\big.\nonumber\\
    &&\quad\quad\quad
    \big.+S^B_{x+1/2,y-1/2}-S^B_{x+1/2,y+1/2}\big]\nonumber\\
    &&\simeq -4J_1\int S^A\partial_x\partial_y S^Bdxdy.
    \label{Eq3}\ea
This term is irrelevant but a perturbative calculation in $J_1$ shows
that two marginal terms are generated at second-order, one of them being
an energy-energy coupling $\varepsilon^A_{x,y}\varepsilon^B_{x,y}$ as in the
Ashkin-Teller model.
\\

The situation was rather clear one decade ago. However, recent
calculations based on Tensor-Network techniques have called these results
into question but have reached opposite conclusions.
A first study based on the Higher-Order Tensor Renormalization Group algorithm
(HoTRG)~\cite{HOTRG} observed varying critical exponents but not in the
Ashkin-Teller universality class~\cite{Li1}. A second study by HoTRG
concluded that the regime of first-order transition is limited to
$1/2<g\lesssim 0.58$, i.e. significantly smaller than previously claimed,
and that the behavior at the tricritical point does not belong to the
4-state Potts universality class, excluding therefore the Ashkin-Teller
universality class along the critical line~\cite{Yoshiyama}.
A recent Monte Carlo simulation yields the same conclusion
on the location of the tricritical point~\cite{Li2}.
The $J_1-J_2$ model was also studied by simulating
the imaginary-time evolution of a Matrix Product State using the iTeBD
algorithm~\cite{Gangat}. It was observed that the first-order regime
extends to $1/2<g<+\infty$, i.e. that the transition becomes continuous
only in the limit $g\rightarrow +\infty$ when $J_1=0$.
\\

The interest in Tensor-Network Renormalization Group techniques
stems from their ability to overcome certain limitations inherent in other
computational methods. Transfer matrix calculations are exact but limited
to small stripes. Monte Carlo simulations, on the other hand, enable the study
of larger systems. However, since there is no cluster algorithm for the
$J_1-J_2$ model, simulations were performed using the Metropolis algorithm
that is known to suffer from a strong critical slowing-down with a dynamical
exponent $z\simeq 2.17$ at the Ising critical point~\cite{Janke}.
Monte Carlo simulations at first-order phase transitions
are also notoriously difficult because the exponentially small probability of
tunneling between the low-temperature phases has to be compensated by
an exponentially large number of Monte Carlo iterations (super critical
slowing-down). Multicanonical algorithms allow to overcome this difficulty.
Tensor-Network techniques open the door to considerably larger lattices.
However, these techniques are variational and come with their own set of
challenges. Whereas statistical errors can be rigorously estimated for
any thermodynamic average computed by Monte Carlo simulation, one can only
check the convergence of the estimates obtained with Tensor-Network algorithms
by performing calculations with different bond dimensions. Errors on
thermodynamic averages cannot be estimated from the truncation error.
Moreover, the convergence of Tensor Renormalization Group algorithms
depends on the particular decomposition of the partition function into
a product of tensors~\cite{Chatelain}.
\\

In this paper, we consider a different $J_1-J_2$ model that is more
suited to simulations by Tensor Renormalization Group algorithms.
The Hamiltonian Eq.~\ref{Eq2} requires to either manipulate rank 8
tensors or to consider rank 4 tensors but with non-independent legs. In
Ref.~\cite{Yoshiyama}, a rank 4 tensor is associated to each plaquette
of the square lattice.
The legs of the tensor do not correspond to the four spins $\sigma_1$,
$\sigma_2$, $\sigma_3$ and $\sigma_4$ at the corners of the plaquette
but to the products $i=\sigma_1\sigma_2$, $j=\sigma_2\sigma_3$,
$k=\sigma_3\sigma_4$, and $l=\sigma_4\sigma_1$ on the four bonds of the
plaquette. The constraint
$ijkl=1$ should always be satisfied. However, after the truncation step
of the HoTRG algorithm, the product $ijkl=1$ is not guarantee to take
exactly the value 1. We consider a different model for which this
potential difficulty does not arise. Moreover, non-local spin-spin
interactions between the two Ising replicas were designed in such a
way to induce a marginal critical behavior according to the mechanism
proposed by Kalz {\sl et al.}~\cite{Kalz1}. It is therefore a good way
to test this mechanism.
\\

In the first section, the model and the BTRG algorithm are presented.
The phase diagram is discussed in the second section. The critical
behavior along the second-order transition line is shown to belong to
the Ashkin-Teller universality class in section III. Conclusions follow.

\section{Model and algorithm}
\subsection{The model and its continuum limit}\label{model}
\begin{figure}
    \centering
    \includegraphics[width=0.27\textwidth]{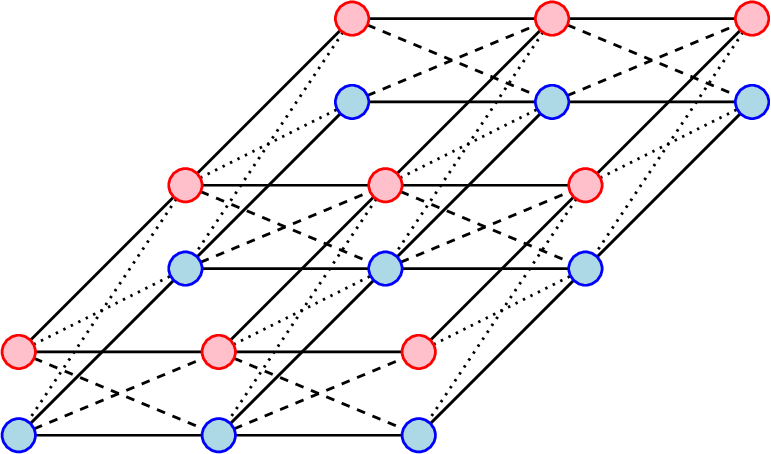}
    \caption{Representation of the model defined by Eq.~\ref{Eq4}.
    For clarity, the two Ising replicas have been shifted in a
    third direction and represented as two layers.
    Red and blue circles correspond respectively to the spins $\sigma^A_{i,j}$
    and $\sigma^B_{i,j}$. The anti-ferromagnetic intra-layer couplings $-J_2$
    are represented as bold lines. The dashed lines are the ferromagnetic
    inter-layer couplings $J_1$ and the dotted ones the anti-ferromagnetic
    inter-layer couplings $-J_1$.}
    \label{fig2}
\end{figure}

We consider two anti-ferromagnetic Ising models $\sigma^A_{i,j}$ and
$\sigma^B_{i,j}$ coupled by a non-local spin-spin interaction.
The Hamiltonian is
    \begin{eqnarray}
    -\beta H&=&-J_2\sum_{i,j}\sigma^A_{i,j}\big[
    \sigma^A_{i+1,j}+\sigma^A_{i,j+1}\big]\nonumber\\
    &&-J_2\sum_{i,j}\sigma^B_{i,j}\big[
    \sigma^B_{i+1,j}+\sigma^B_{i,j+1}\big]\nonumber\\
    &&+J_1\sum_{i,j}\sigma^A_{i,j}\big[
    \sigma^B_{i+1,j}-\sigma^B_{i,j+1}\big]
    +A\leftrightarrow B
    \label{Eq4}
    \end{eqnarray}
The first two lines correspond to the Hamiltonian of two replicas
of an anti-ferromagnetic Ising model on a square lattice. For simplicity,
the two replicas lay on the same lattice. One of the two replicas can be shifted
by half a lattice step in both $i$ and $j$ directions to get something closer
to Eq.~\ref{Eq1}. The discussion that follows would not be changed. The third
line of Eq.~\ref{Eq4} is an interaction term that couples these two replicas.
The interaction is ferromagnetic on horizontal edges and anti-ferromagnetic
on vertical ones. $A\leftrightarrow B$ means that the term should be
repeated after the exchange of $A$ and $B$. The model is depicted on
Fig.~\ref{fig2}. The phase diagram of this model
is expected to display more symmetries than the original $J_1-J_2$ model
defined by Eq.~\ref{Eq1}. It is indeed symmetric under
a change of sign of $J_1$ since the Hamiltonian is invariant under the
transformation
    \be J_1\rightarrow -J_1,\quad\quad
    \sigma^B_{i,j}\rightarrow -\sigma^B_{i,j}.\label{ExchJ1-J1}\ee
which amounts to a simple rotation by $90^\circ$ of the lattice.
It is also symmetric under the exchange of $J_1$ and $J_2$. The
Hamiltonian is indeed invariant under $J_1\leftrightarrow J_2$ if
the spins of the two replicas are first exchanged on every two sites
    \be \big(\sigma^A_{i,j},\sigma^B_{i,j}\big)
    \rightarrow\left\{\begin{aligned}
    &\big(\sigma^A_{i,j},\sigma^B_{i,j}\big),\quad i+j {\rm\ even},\\
    &\big(\sigma^B_{i,j},\sigma^A_{i,j}\big),\quad i+j {\rm\ odd}
    \end{aligned}\right.        \label{ExchJ1J2}\ee
and then flipped on every two lines
    \be \sigma^{A,B}_{i,j}\rightarrow (-1)^i\sigma^{A,B}_{i,j}.\label{ExchJ1J2b}\ee

In the following, we will show that the Hamiltonian Eq.~\ref{Eq4} leads to
the same continuum limit as Eq.~\ref{Eq2}. The first step is to rewrite the
inter-layer Hamiltonian (third and fourth lines of Eq.~\ref{Eq4})
as a sum over the plaquettes of a checkerboard lattice~:
    \be J_1\!\!\sum_{i,j,\atop i+j\ {\rm even}}\!\!
    \big(\sigma^A_{i,j}-\sigma^A_{i+1,j+1}\big)
    \big(\sigma^B_{i+1,j}-\sigma^B_{i,j+1}\big)+A\leftrightarrow B\ee
Introducing the staggered magnetization $S^A_{i,j}=(-1)^{i+j}\sigma^A_{i,j}$,
as done by Kaltz {\sl et al.}, only changes $J_1$ into $-J_1$. As discussed
above, this does not affect the phase diagram of the model. Performing now
a $45^\circ$ rotation of the lattice around the center of the plaquette,
the interaction Hamiltonian becomes
    \ba &&-J_1\sum_{i,j}\big(S^A_{x-1/2,y}-S^A_{x+1/2,y}\big)
    \big(S^B_{x,y+1/2}-S^B_{x,j-1/2}\big)\nonumber\\
    &&\simeq J_1\int \partial_x S^A\partial_y S^B\ \!dxdy.
    \ea
After an integration by part, a term similar to Eq.~\ref{Eq3} is obtained.

\subsection{The BTRG algorithm}\label{btrg}
\begin{figure}
    \centering
    \includegraphics[width=0.47\textwidth]{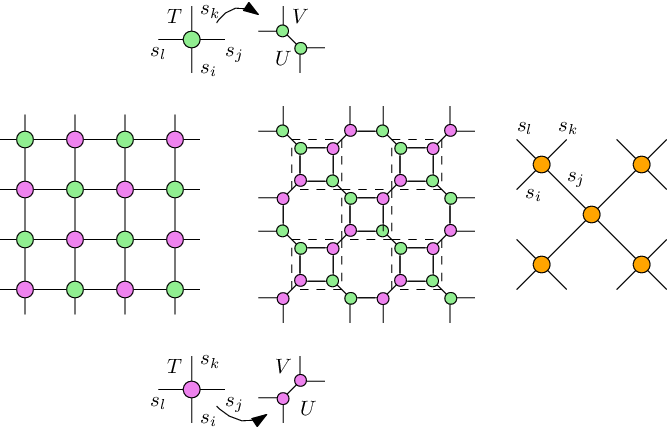}
    \caption{TRG algorithm on the square lattice. Tensors are represented
    as disks with different colors on even and odd sites. The bonds between
    them carry the variables $s_i$ that have to be integrated out to
    compute the partition function. Each variable $s_i$ appears among the
    indices of the two tensors at both edges of the bond $i$. Above and below,
    the Singular Value Decomposition of the tensor $T$ on odd and even sites allows
    to write the tensor $T$ as the contraction of two rank 3 tensors
    $U\sqrt\Lambda$ and $\sqrt\Lambda V$.
    In the center, the new lattice obtained after the decomposition of all tensors
    $T$. On the right, the bonds of the square plaquettes have been integrated
    out, leaving a square lattice of new effective tensors represented
    in orange color.}
    \label{fig3}
\end{figure}

The BTRG algorithm derives from the Tensor Renormalization Group (TRG) algorithm
introduced by Levin and Nave~\cite{Levin,Xiang}. The original TRG algorithm was
proposed first for the triangular lattice before being extended to the square
lattice~\cite{Gu}. The starting point is a decomposition of the partition
function as a product of rank 4 tensors:
    \be {\cal Z}=\sum_{s_1,s_2,\ldots} \prod_{\alpha\in V,\atop
    i,j,k,l\in E_{\alpha}} T_{s_is_js_ks_l}.    \label{Eq5}\ee
A tensor $T$ is located at each vertex $\alpha$ of the lattice. $E_{\alpha}$
denotes the subset of edges of the lattice connecting the vertex $\alpha$
to its neighbors. Bond variables $s_i$ are carried by the edges of the lattice.
The bond variable $s_i$ appears among the indices of the two tensors located
at the vertices that are connected by the bond $i$. In the simple case of the
Ising model, a possible decomposition of the partition function consists in
identifying the bond variables $s_i$ with the Ising spins $\sigma_i$
(or with $(\sigma_i+3)/2\in\{1,2\}$ in the numerical implementation).
The tensors $T$ are then located at the center of the plaquettes of the
square lattice and correspond to the Boltzmann weight
of this plaquette:
    \be T_{\sigma_i\sigma_j\sigma_k\sigma_l}
    =e^{\beta J(\sigma_i\sigma_j+\sigma_j\sigma_k+\sigma_k\sigma_l
    +\sigma_l\sigma_i)}.\ee
This formulation makes it easy to construct the effective statistical weight
$T^{\rm eff}$ resulting from the decimation of a spin $\sigma_i$. $T^{\rm eff}$
is given by the contraction of the two tensors at the two edges of the bond
carrying the spin $\sigma_i$
    \be  T^{\rm eff}_{\sigma_j\sigma_k\sigma_l\sigma_j'\sigma_k'\sigma_l'}
    =\sum_{\sigma_i=\pm 1}T_{\sigma_i\sigma_j\sigma_k\sigma_l}
    T_{\sigma_i\sigma_j'\sigma_k'\sigma_l'}.\ee
It is a rank 6 tensor. Iterating this procedure leads to a partition function with
fewer and fewer spins but with tensors of larger and larger ranks that rapidly
become unmanageable by a computer. The solution proposed by Levin and Nave relies
on a Singular Value Decomposition (SVD) of each tensor before any decimation.
The SVD is performed in two different ways:
        \be T_{s_is_js_ks_l}=\left\{\begin{aligned}
        &\sum_{s_n} U_{s_is_j;s_n}\Lambda_{s_n}V_{s_ks_l;s_n},
        \quad {\rm even\ sites},\\
        &\sum_{s_n} U_{s_ls_i;s_n}\Lambda_{s_n}V_{s_js_k;s_n}
        \quad {\rm odd\ sites},
        \end{aligned}\right.\label{SVD}\ee
for odd and even lattice sites (Fig.~\ref{fig3}).
$U$ and $V$ are unitary matrices that are
reshaped into rank 3 tensors. Each rank 4 tensor $T$ is replaced by the
contraction of either the two rank 3 tensors $U_{s_is_js_n}\sqrt{\Lambda_{s_n}}$
and $\sqrt{\Lambda_{s_n}}V_{s_ks_ls_n}$ on even sites or $U_{s_ls_is_n}
\sqrt{\Lambda_{s_n}}$ and $\sqrt{\Lambda_{s_n}}V_{s_js_ks_n}$ on odd sites.
As can be seen on Fig.~\ref{fig3}, the new tensors form a lattice with two
kinds of plaquettes with either 4 or 8 sites. The 4 tensors of the plaquettes
with 4 sites can be contracted, leaving a single rank 4 tensor. Performing
this operation in all such plaquettes leads to a new lattice of rank 4 tensors
with a lattice step larger by a factor $\sqrt 2$ and oriented at $45^\circ$
of the initial one. The above-detailed procedure is exact. However, if the
dimension of the tensors is initially $\chi^4$, the dimension of the
rank 3 tensors after the SVD of Eq.~\ref{SVD} is $\chi\times\chi\times\chi^2$.
After contraction, the final tensors have a dimension $(\chi^2)^4$. Again,
after a few iterations of this algorithm, the tensors become exponentially
large. To circumvent this problem, the SVD Eq.~\ref{SVD} can be limited
to the $\chi$ largest singular values. This truncation ensures that the
dimension of the rank 3 tensors $U$ and $V$ are $\chi^3$, leading to
a tensor $T$ of dimension $\chi^4$ after contraction.
This algorithm belongs to the class of variational methods. Indeed,
the truncation minimizes the error defined as the Hilbert-Schmidt
norm $||T-T_{\rm trunc}||$ where $T_{\rm trunc}$ is the result of the
truncation of $T$.
\\

\begin{figure}
    \centering
    \includegraphics[width=0.47\textwidth]{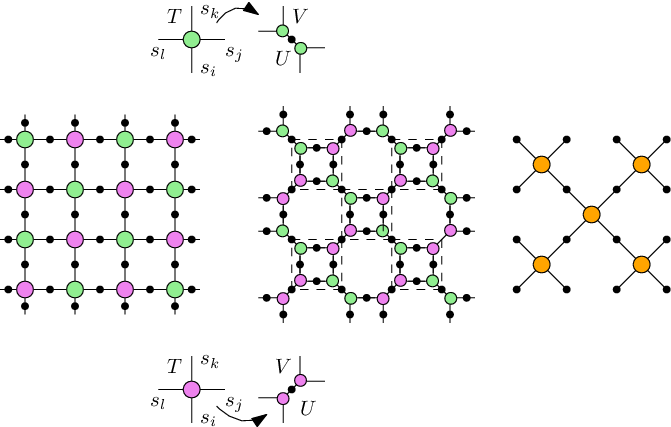}
    \caption{BTRG algorithm on the square lattice. The difference with the
    TRG algorithm lies in the presence of diagonal tensors on each edge
    of the lattice. They are represented as black dots on the figure.}
    \label{fig4}
\end{figure}

The accuracy of the TRG algorithm can be improved by introducing a weight
$w$ on each bond of the lattice~\cite{BTRG}. The decomposition Eq.~\ref{Eq5}
of the partition function is replaced by
    \be {\cal Z}=\sum_{s_1,s_2,\ldots} \prod_{\alpha\in V,\atop
    i,j,k,l\in E_{\alpha}} T_{s_is_js_ks_l}\ \!\prod_i \omega_i.
    \label{Eq6}\ee
The Bond-Weighted Tensor Renormalization Group algorithm (BTRG) is very
similar to the TRG algorithm. The main difference lies in the way the rank 4
tensors $T$ are decomposed after the SVD of Eq.~\ref{SVD}. Each tensor $T$
is replaced by the contraction of either the two rank 3 tensors
$U_{s_is_js_n}(\Lambda_{s_n})^k$ and $(\Lambda_{s_n})^kV_{s_ks_ls_n}$
on even sites or $U_{s_ls_is_n}(\Lambda_{s_n})^k$ and $(\Lambda_{s_n})^{k}
V_{s_js_ks_n}$ on odd sites. Between these two rank 3 tensors, a new weight
$(\Lambda_{s_n})^{1-2k}$ is introduced. $k$ is a free parameter. The TRG
algorithm is recovered when $k=1/2$. It was suggested that the optimal
choice is $k=-1/2$~\cite{BTRG}. The algorithm is depicted on Fig.~\ref{fig4}.

Free energy and critical exponents can be estimated from
the $\chi\times\chi$ transfer matrix $M$ of the system obtained by contraction
of the tensor $T$~\cite{Ueda,Ueda2,Huang,Guo}:
    \be M_{s_i,s_k}=\sum_{s_j}T_{s_is_js_ks_j}.\ee
The largest eigenvalues $\lambda_i$ of the transfer matrix $M$ are then
estimated using the Lanczos algorithm as implemented in the {\tt Arpack}
library~\cite{arpack}. The free energy density of the system is given by
the logarithm of the largest eigenvalue $\lambda_0$:
    \be f(L)=-{1\over L}\ln\lambda_0        \label{FSSf}\ee
where the width $L$ of the system is related to the number $n$
of BTRG iterations by $\sqrt 2^n$.

\section{Phase diagram}
\subsection{Phases and transitions}\label{Phases}
\begin{figure}
    \centering
    \includegraphics[width=0.47\textwidth]{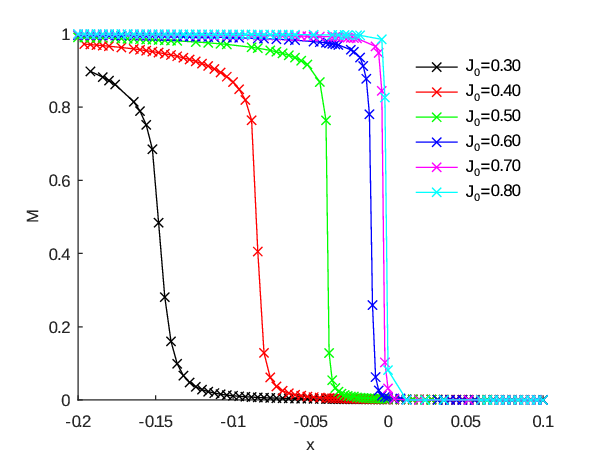}
    \caption{Average staggered magnetization density $\langle m^A\rangle$
    along lines perpendicular to the diagonal $J_1=J_2$.
    The staggered magnetization has been estimated by coupling
    a small magnetic field $h=10^{-3}$ to the system.}
    \label{fig10}
\end{figure}

The state of the system is readily determined at several points of the
phase diagram. At the point $J_1=J_2=0$, equivalent to an infinite
temperature, the Ising spins are uncoupled so the equilibrium state
of the system is the paramagnetic state. One may therefore assume that
there exists a finite region of the phase diagram, containing the point
$J_1=J_2=0$, where the paramagnetic phase is stable.
In the limit $J_2\rightarrow +\infty$ and $J_1\rightarrow 0$, the
two replicas are uncoupled and ordered anti-ferromagnetically.
An anti-ferromagnetic phase is therefore expected in a region
the phase diagram containing this point. The two average staggered
magnetizations
    \be \langle M^{A,B}\rangle=\sum_{i,j} (-1)^{i+j}
    \langle\sigma^{A,B}_{i,j}\rangle    \label{StaggM}\ee
are expected to take a non-zero value in this phase when a small
magnetic field $h$ is coupled to the system by a Zeeman Hamiltonian
$hM^{A,B}$. We measured these staggered magnetizations along different
lines perpendicular to the diagonal $J_1=J_2$ and parameterized as
    \be J_1(x)=J_0+x,\quad\quad J_2(x)=J_0-x\ee
where the parameter $x$ allows to move along these lines and $J_0$
identifies the different lines by their intersection $J_1=J_2=J_0$
with the diagonal.
BTRG simulations were performed with $\chi=32$ states and 32 iterations
for several values of the staggered magnetic field $h$. The staggered
magnetization density is estimated from the finite-difference
of the free energy density
    \be\langle m^A\rangle=-\left({\partial f\over\partial h}
    \right)_{h\rightarrow 0^+}\simeq -{f(h)-f(0)\over h}.\ee
A stable estimate is obtained for small magnetic fields $h\simeq 10^{-3}$.
A phase transition is clearly observed on Fig.~\ref{fig10}. The transition
becomes steeper as $J_0$ is increased. This suggests the possibility of a
first-order phase at large $J_0$.

In the limit $J_1\rightarrow +\infty$ and $J_2\rightarrow 0$,
the Ising spins are only coupled to their neighbors in the other
replica. The coupling is ferromagnetic in the horizontal direction
of the lattice but anti-ferromagnetic in the vertical one.
The quantity
    \be \bar M^A=\sum_{i,j\atop{\rm even}}\big[\sigma^A_{i,j}
    +\sigma^B_{i+1,j}-\sigma^A_{i+1,j+1}-\sigma^B_{i,j+1}\big]
    \label{ReplMagn}\ee
is non-zero on average when the up-down symmetry is broken by a magnetic
field coupled to $\bar M^A$. Unfortunately, it is not possible to add this
Zeeman coupling to our implementation of the BTRG algorithm because we
assumed that all vertices of the tensor network were equivalent, while
such a Zeeman coupling implies inequivalent plaquettes. Nevertheless,
the behavior of $\langle\bar M^A\rangle$ can be deduced from the symmetry
of the Hamiltonian under the exchange $J_1\leftrightarrow J_2$ when the
transformations Eq.~\ref{ExchJ1J2} and \ref{ExchJ1J2b} are successively
performed. One can check that the image of the order parameter
Eq.~\ref{ReplMagn} under these transformations is, as expected, the
staggered magnetization Eq.~\ref{StaggM}. As consequence, the behavior
of $\langle\bar M^A(x)\rangle$ along the lines perpendicular to the diagonal
$J_1=J_2$ is simply given by the reflection $\langle M^A(-x)\rangle$
of the staggered magnetization.

\subsection{Critical lines}
As discussed above, the behavior of the staggered magnetization shows
the existence of a transition line in the half plane $J_2>J_1$ of the phase
diagram with possibly a first-order regime. Because of the symmetry of the
model under the exchange $J_1\leftrightarrow J_2$, the same transition line
is expected in the half-plane $J_1>J_2$. In this section, we determine more
precisely the location of these two transition lines in the second-order regime
and study the critical behavior.
\\

\begin{figure}
    \centering
    \includegraphics[width=0.47\textwidth]{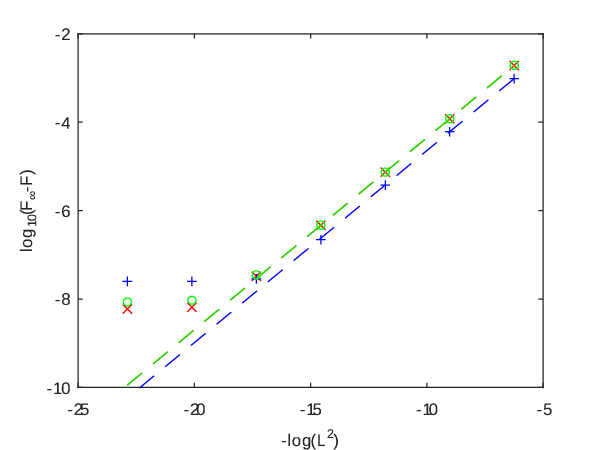}
    \caption{Logarithm (base 10) of the free energy difference
    $F_\infty-F(L)$ versus the logarithm of $1/L^2$ for couplings $(J_1,J_2)$
    equal to $(0.612965, -0.605170)$, $(0.448669, -0.134480)$, and
    $(0.440618, 0)$. The symbols correspond to the data points and the dashed
    curves to the quadratic fit. The central charge $c$ is related to the vertical
    intercept. The number of states is $\chi=48$.}
    \label{fig8}
\end{figure}

Assuming that conformal invariance holds for this system, the free energy
density is expected to scale with the stripe width $L$ as~\cite{Cardy}
    \be f(L)=f_\infty-{\pi c\over 6L^2}+{\cal O}\bigg({1\over L^4}\bigg)
    \label{FSS-f}\ee
at the critical point. The universal constant $c$ is the central charge
which takes the value $c=1/2$ for the 2D Ising model and $c=1$ for the
Ashkin-Teller model. Away from criticality, the constant $c$ is not universal
anymore and takes a smaller value. However, it was shown to that this constant
increases monotonically along the Renormalization-Group flow and is maximum at
the fixed point~\cite{Zamo}. The critical point can therefore be determined as
the location of the maximum of $c$. We estimated the free energy density
from the largest eigenvalue of the transfer matrix (Eq.~\ref{FSSf}). The
central charge $c$ is then estimated from the Finite-Size Scaling Eq.~\ref{FSS-f}.
To take into account the first correction to this behavior,
a quadratic fit with $1/L^2$ was performed:
    \be f(L)=f_\infty-{\pi c\over 6L^2}+{a\over L^4}.\ee
A cubic fit does not yield significantly different results. Two difficulties
were however encountered: the lattice size is multiplied by a factor $\sqrt 2$
at each iteration of the BTRG algorithm. As a consequence, our lattice
sizes are distributed exponentially, and not linearly, as would be the
case with a more traditional transfer matrix calculation. We have checked
for the Ising model and the Ashkin-Teller model along its critical line
that the fit gives nevertheless the expected central charge. We assume that
it is also the case for our $J_1-J_2$ model. The second difficulty is that
the free energy does not behave as $1/L^2$ at large lattice
sizes (our largest lattice size is $\sqrt 2^{33}\simeq 92,680$) but tends
towards a plateau, due to the finite number of states $\chi$ kept in
the BTRG calculation and to the finite accuracy in the estimation of the
largest eigenvalue $\lambda_0$ with the {\tt Arpack} library.
To circumvent this problem, we discarded all free energies $f(L_n)$
such that $|f(L_n)-f(L_{n-1})|<10^{-12}$ where $\{L_n\}$ are the set
of lattices sizes given by the BTRG algorithm. Examples of fits
of the free energy density are shown on Fig.~\ref{fig8} for three different
points on the critical line. Because the lattice sizes are distributed
exponentially, logarithms are plotted. One clearly sees on Fig.~\ref{fig8}
the plateau reached by the free energy at large lattice sizes.

\begin{figure}
    \centering
    \includegraphics[width=0.47\textwidth]{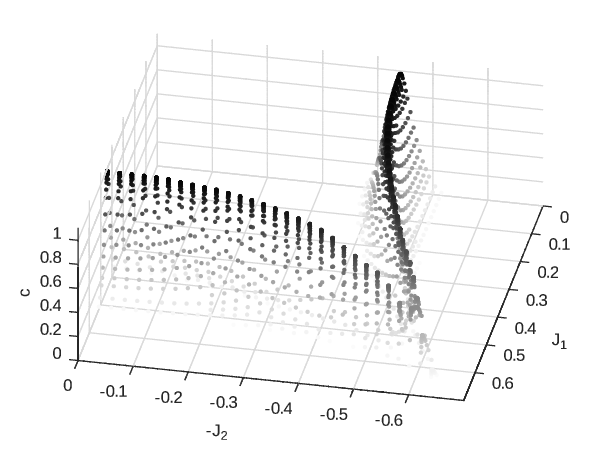}
    \caption{Central charge $c$ versus the couplings $J_1$ and $-J_2$
    for $\chi=16$ states in the BTRG algorithm.}
    \label{fig5}
\end{figure}

The critical lines of the phase diagram have been determined by searching
for the maximum of the central charge $c$.
A rough estimate of the location of the critical lines was obtained by
performing a scan of the plane $(J_1,J_2)$ while keeping only
$\chi=16$ states in the BTRG algorithm. The location of the maxima was
then refined by dichotomy until reaching an accuracy of $10^{-5}$.
A few points around the maxima were selected and used to initiate
a new search by dichotomy with $\chi=24$ states. The procedure was
repeated for $32$ and $48$ states.
The central charge is plotted for $\chi=16$ in Fig.~\ref{fig5}.
Two branches are clearly observed. As expected (Sec.~\ref{model}),
they are images of each other under the exchange $J_1\leftrightarrow J_2$.
The two branches seem to merge at $J_1=J_2\simeq 0.6$. Beyond this point,
i.e. for $J_1\gtrsim 0.6$, the free energy density $f(L)$ reaches a plateau
already for small lattice sizes $L$ so that no fit can be performed.
For $\chi=24$, the free energy density $f(L)$ can be fitted only
for $J_1\lesssim 0.40$ (first branch) or $J_2\lesssim 0.40$ (second branch).
For $\chi=32$ and 48, the fit is reliable only for $J_1\lesssim 0.33$
(first branch) or $J_2\lesssim 0.33$ (second branch). In contrast to
the case $\chi=16$, the two branches do not merge anymore. This does not
imply that there is no phase transition for $J_1\gtrsim 0.33$.
A phase transition was indeed observed in Sec.~\ref{Phases}.
However, the relation Eq.~\ref{FSSf} holds only for second-order phase
transitions when the RG fixed point is conformally invariant.
The phase transition beyond $J_1\gtrsim 0.33$ is therefore probably of
first order.

\begin{figure}
    \centering
    \includegraphics[width=0.47\textwidth]{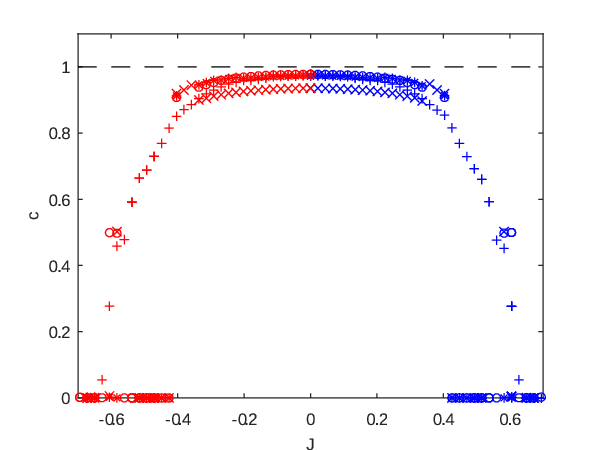}
    \caption{Maximum of the central charge $c$ in the two branches
    versus the coupling $J_1$ in the first branch (blue symbols)
    and versus $-J_2$ in the second branch (red symbols).
    The different symbols correspond to different number of states
    ($\chi=16$ for $+$, 24 for $\times$, 32 for $\ast$ and 48 for $\circ$).}
    \label{fig6}
\end{figure}

The maxima of the central charge are plotted in the two branches on
Fig.~\ref{fig6}. For $\chi=16$, the central charge is approximatively
constant for $J_1,J_2\lesssim 0.4$, then decreases rapidly and
vanishes for $J_1,J_2\simeq 0.6$. For $\chi=24,32$ and $48$, this
decrease is much less pronounced. Instead, the central charge varies
slowly from $c=0.9774$ at $(J_1,J_2)$ equal to $(0,-0.44062)$ and
$(0.44062, 0)$~\footnote{Note1} to $c=0.9383$ at $(J_1,J_2)$ equal to
$(0.33620,-0.49206)$ and $(0.49206,-0.33620)$ for $\chi=48$. Close values
are obtained for $\chi=32$ while much larger deviations are observed for
$\chi=24$. Surprisingly, a few points at $c\simeq 0.498$ around $(J_1,J_2)
\simeq (0.61,-0.59)$ and $(0.59,-0.61)$ can be seen on the figure for $\chi=48$
but not for the other number of states. The free energy density of one of
these points is plotted on Fig.~\ref{fig8}. Nothing special can be
observed. Our interpretation is that, probably due to a numerical instability,
the BTRG algorithm took the system to a fixed point which is critical
for one of the two replicas but trivial (either paramagnetic or
anti-ferromagnetic) for the other.

In the Ashkin-Teller model, the central charge is expected to remain constant
at $c=1$ along the self-dual critical line. Our numerical data for the $J_1-J_2$
Ising model show deviations that are at most $6\%$ from this value.

\section{Critical behavior}
\begin{figure}
    \centering
    \includegraphics[width=0.47\textwidth]{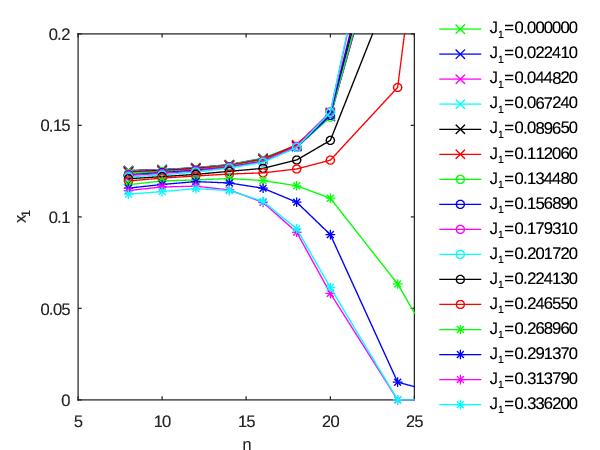}
    \caption{First scaling dimension $x_1$ versus the number of iterations
    of the BTRG algorithm with $\chi=96$. The different curves correspond
    to different points on the first critical line.}
    \label{fig9}
\end{figure}

The scaling dimensions $x_n$ of the scaling operators of the theory can be
estimated from the gaps between the eigenvalues $\lambda_n$ of the transfer
matrix and the largest one $\lambda_0$. The gap-exponent relation states that
    \be x_n=-{L\over 2\pi}\ln{\lambda_n\over\lambda_0}.
    \label{GapExposant}\ee
To improve the accuracy on the $x_n$, additional calculations were performed
with $\chi=64$ and 96 states along the two critical lines previously determined
with $\chi=48$. As discussed in details in Ref.~\cite{Ueda2}, the estimation
of the scaling dimensions require to choose carefully the lattice size.
At too small lattice sizes,
Finite-Size corrections cannot be neglected and yield systematic
deviations of the estimated scaling dimensions. At intermediate lattice
sizes, a plateau is observed on Fig~\ref{fig9} for various points on the
critical line. However, at large lattice sizes, the estimates of the
scaling dimensions either diverge or tends to zero, as would be case
in the ferromagnetic or paramagnetic case. Due to the finite number of
states $\chi$ kept in the BTRG algorithm, the system is indeed gapped
and not really critical. We measured the scaling dimensions in the plateau,
after 12 iterations of the BTRG algorithm.

\begin{figure}
    \centering
    \includegraphics[width=0.47\textwidth]{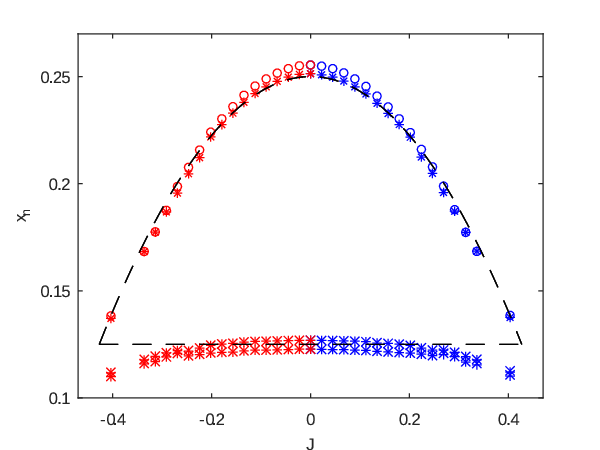}
    \caption{Three first scaling dimensions $x_n$ along the two critical
    lines versus the coupling $J_1$ in the first branch (blue symbols)
    and versus $-J_2$ in the second branch (red symbols). The different
    symbols correspond to $n=1$ for $+$, 2 for $\times$ and 3 for $\circ$.
    The black dashed curves are only guide to the eyes.}
    \label{fig7}
\end{figure}

The three first scaling dimensions along the two critical curves are
plotted on Fig.~\ref{fig7}. The difference between the first two
scaling dimensions $x_1$ and $x_2$ is at most $4.10^{-4}$ so it can be
assumed that they are degenerated. They take a value close to $1/8$ at the
points $(0,{1\over 2}\ln(1+\sqrt 2))$ and $(-{1\over 2}\ln(1+\sqrt 2),0)$
where the two Ising replicas are not coupled. When going away from these
points, they decrease down to a value $0.118$ at $\chi=64$ and $0.116$
at $\chi=96$, i.e. a relative deviation of $6\%$ from $1/8$. Note that
the same decrease was observed for the central charge. Assuming that $x_1$
and $x_2$ are constant and equal to $x_{\sigma}=1/8$, one can associate
them to the magnetization density of each Ising replicas.
\\

As can be observed on Fig.~\ref{fig7}, the third scaling dimension $x_3$
takes the value $0.254$ at $\chi=64$ and 0.255 at $\chi=96$, close to $1/4$,
at the points where the Ising replicas decouple and decreases significantly
along the two critical lines. Remarkably, the data points fall reasonably
close to a simple parabola ${1\over 4}-{1\over 8}(J_1/J_t)^2$ with
$J_t\simeq 0.427$. The range of variation of the third scaling dimension $x_3$
is similar to that of the scaling dimension $x_{\sigma\tau}$ of the polarization density
of the Ashkin-Teller model, which decreases along the self-dual critical line,
going from $1/4$ at the Ising point to $1/8$ at the 4-state Potts point.
If the similarity is actually a correspondence, it would imply that our
$J_1-J_2$ model belongs to the 4-state Potts universality class when
$J_1=J_t$ (resp. $J_2=J_t$) on the first (resp. second) branch.

\begin{figure}
    \centering
    \includegraphics[width=0.47\textwidth]{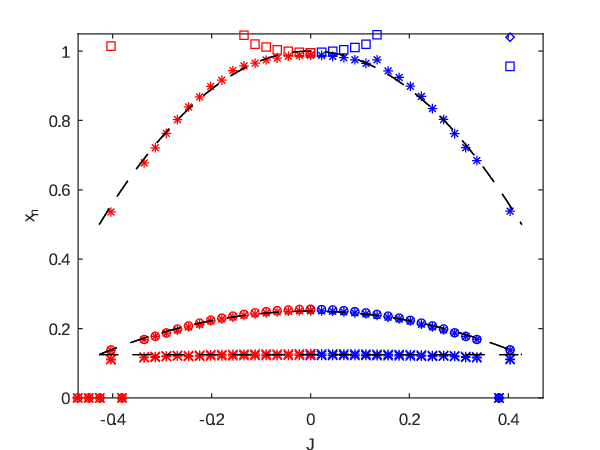}
    \caption{Five first scaling dimensions $x_n$ along the two critical
    lines versus the coupling $J_1$ in the first branch (blue symbols)
    and versus $-J_2$ in the second branch (red symbols).
    The black dashed curves are only guide to the eyes.
    The dashed curve for $x_4$ ($\ast$) has been computed using a parabolic
    approximation of  $x_3$ ($\circ$) and the assumption of a Ashkin-Teller
    universality class.}
    \label{fig11}
\end{figure}

The fact that the third scaling dimension goes from $1/4$ to $1/8$, like
the scaling dimension $x_{\sigma\tau}$ of the polarization density of the
Ashkin-Teller model, is not sufficient to declare that our $J_1-J_2$ model
belongs to the Ashkin-Teller universality class. In the latter, the critical
exponents have been shown to be~\cite{Kadanoff,Nienhuis,Baxter}
    \be x_{\sigma}={1\over 8},\quad
    x_{\sigma\tau}={1\over 8-4y},\quad
    y_t={3-2y\over 2-y}         \label{ScalDimAT}\ee
where the parameter $y$ is in the range $[0;3/2]$ along the critical line.
Using the ansatz $x_{\sigma\tau}={1\over 4}-{1\over 8}(J_1/J_t).^2$ that
was introduced above, we extracted the parameter $y$ and plotted the scaling
dimension $x_t=2-y_t$ of the energy density. As can be observed on
Fig.~\ref{fig11}, the dependence on $J_1$ (resp. $J_2$) of the 4th scaling
dimensions of our $J_1-J_2$ model is in good agreement with this prediction
of $x_t$. This provides strong evidence that the model belongs to the
Ashkin-Teller universality class.
\\

One can see on the figure that the gap with the 5th scaling dimension
vanishes at the Ising point for $\chi=96$. It is a pure coincidence.
For $\chi=64$, the 5th scaling dimension $x_5$ takes a value close to
$\simeq 0.92$. A crossing with the 4th scaling dimensions is therefore
observed for $J_1\simeq 0.13$. For $\chi=48$, $x_5\simeq 0.82$ and the
crossing with $x_4$ occurs at larger couplings $J_1$. We therefore
expect that, for $\chi>96$, the 5th scaling dimension take values larger
than 1 so that the gap with $x_4$ does not close anymore.
One should also mention that, as discussed earlier, the estimation
of the critical line is reliable only for $J_1\lesssim 0.33$ or
$J_2\lesssim 0.33$. As a consequence, all scaling dimensions $x_n$
computed for $J_1>0.33$ or $J_2>0.33$ take values either very small or
very large and, in this case, are not visible on the figure. Surprisingly,
the two points at $J_1\simeq 0.40344$ and $J_2\simeq 0.40344$ are an
exception and lead to estimates of the scaling dimensions $x_n$ consistent
with the Ashkin-Teller universality class.

\section{Conclusions}
In this study, we have explored the phase diagram of a new 2D frustrated
Ising model with non-local spin-spin interactions sharing the same continuum
limit as the $J_1-J_2$ model. Using the BTRG Tensor-Network algorithm, we
have provided evidence that the two transition lines, related by the symmetry
$J_1\leftrightarrow J_2$, include both a first and a second-order regime.
Even though our model is identical to the $J_1-J_2$ model only in the scaling
limit, implying that only universal quantities are expected to match in the
two models, our conclusions tend to be in line with most studies on the
$J_1-J_2$ model and contradict the iTeBD study that concluded to
a second-order regime limited to the point $J_1=0$ of the phase
diagram~\cite{Gangat}.
\\

In the second-order regime, our estimates of the central charge and of the
magnetic, electric, and thermal critical exponents along the critical line
are compatible with the Ashkin-Teller universality class. This result is in
agreement with the Monte Carlo simulations performed one decade
ago~\cite{Kalz1,Jin1,Kalz2,Jin2} but not with the more recent Tensor-Network
calculations~\cite{Li1,Yoshiyama}. This also demonstrates the validity of
the analysis of the scaling limit of the $J_1-J_2$ model reported in
Ref.~\cite{Kalz1}.

\section*{Acknowledgments}
This work was supported by the french ANR-PRME UNIOPEN grant (ANR-22-CE30-0004-01).

\end{document}